# *Temporal metamaterials with gain and loss*

*Victor Pacheco-Peña[1] and Nader Engheta[2]*

[1]*School of Mathematics, Statistics and Physics, Newcastle University, Newcastle Upon Tyne, NE1 7RU, United Kingdom*
[2]*Department of Electrical and Systems Engineering, University of Pennsylvania, Philadelphia, PA 19104, USA*
*email: Victor.Pacheco-Pena@newcastle.ac.uk, engheta@ee.upenn.edu

**Manipulation of wave-matter interactions in systems with loss and gain have opened new mechanisms to control wave propagation at will. Metamaterials and metasurfaces having spatially inhomogeneous loss and gain have been studied in the past few years by exploiting parity-time (PT) symmetry concepts inspired from quantum-physics. In this work we theoretically study the control of light-matter interactions in spatially unbounded metamaterials having a time-modulated permittivity whose imaginary part is temporally modulated to induce loss and gain, while the real part stays unchanged. We show both numerically and theoretically how such temporally modulated multistepped metamaterials with loss and gain can be equivalent to a temporal effective metamaterial having an effective permittivity modelled by a step function in time. Interestingly, it is shown how the amplitude of a monochromatic electromagnetic wave traveling inside such temporal metamaterials can experience spatiotemporal decay or amplification depending on the values of loss and gain added into the system, while its wavenumber is preserved. We envision that our findings may open new avenues in exploration of potential applications of temporal metamaterials in signal amplification and loss mitigation.**

The control and manipulation of wave propagation has been a key to the current technological advancements. For instance, the fields of optics, photonics and plasmonics have been devoted to studying light-mater interactions in order to provide a full physical understanding on how light can be manipulated at will[1]. Another paradigm for controlling and tailoring waves and fields is artificially engineered materials (metamaterials)[2]. Metamaterials (and metasurfaces as their two-dimensional (2D) version) have opened new avenues in the design of artificial media with extreme electromagnetic (EM) responses (permittivity, ε, permeability, µ) such as negative refraction [3–5] and epsilon-near-zero[6–9]. Metamaterials and metasurfaces have been demonstrated in various spectral ranges from the radio frequencies to the optical regime[10–12], allowing their implementation in improved and new applications such as sensing[13–15], computing with acoustics and EM waves[10,16,17], lensing[18–22] and designs using deep learning techniques [23–25], to name a few.

As most of these applications are designed using passive artificial media, unavoidable losses can affect their overall performance[26]. However, losses are not always a drawback in light-matter interactions with passive media as metamaterials can be tailored to manage their effect. In this realm, a prominent research area that has taken off within the past decade involves the manipulation of EM wave propagation via metamaterials and metasurfaces having spatially modulated/inhomogeneous loss and gain[27–29]. Motivated by systems with parity-time symmetry (*PT*), a concept first developed in quantum-physics[28,30,31], metamaterials exhibiting spatially modulated loss and gain have enabled new sets of intriguing and exotic physical phenomena such as unidirectionality [32], extreme anisotropy[33], negative refraction[34] and invisibility cloaking[35].

All these applications have been mostly studied in the frequency domain (time-harmonic scenario) with spatial inhomogeneity in material parameters. However, spatiotemporal metamaterials are becoming a new paradigm with renewed interest for controlling fields and waves in both space and time, i.e., in four dimensions 4D [36,37]. The study of EM wave propagation within time-modulated media was first studied in the last century [38,39] where the propagation of a monochromatic EM wave is explored theoretically inside a material having a temporal relative permittivity ($\varepsilon_r$) rapidly changed in time, $\varepsilon_r(t)$ (with a rise/fall time smaller than the period $T$ of the wave inside the medium) from $\varepsilon_{r1}$ to $\varepsilon_{r2}$ at a time t = $t_1$ (with real-valued $\varepsilon_{r1}, \varepsilon_{r2} \geq 1$). In that analysis it was shown that such rapid change in $\varepsilon_r$ can generate a temporal boundary, as the temporal analogue of a spatial boundary between two media, which leads to a forward (FW) and a backward (BW) wave. Temporal and spatiotemporally modulated media have offered new paradigms for exciting light-matter interactions enabling their proposal in groundbreaking applications[40–42] in the time domain such as Fresnel drag [43], nonreciprocity [44,45], inverse prism [46], antireflection temporal coatings [47,48], meta-atoms [49,50], holography and frequency conversion [37,51], temporal aiming [52] and temporal Brewster angle [53].

Motivated by the opportunities of spatiotemporal metamaterials and the interesting physical phenomena arising from wave propagation in metastructures with *PT* symmetry, in this Letter we propose and explore metamaterials with time-varying permittivity functions $\varepsilon_r(t)$ with temporally modulated gain and loss. In a recent work, we have shown how effective medium concepts known in the frequency domain such as spatial multilayered metamaterials can have their temporal analogue in temporal metamaterials whose $\varepsilon_r$ is changed periodically in time between $\varepsilon_{r1}$ and $\varepsilon_{r2}$ (all values larger than 1) [54]. In so doing, a temporal effective permittivity $\varepsilon_{reff}$ can be achieved. In the present work, we build upon this to explore EM wave propagation in temporal metamaterials having temporally periodic loss and gain. We provide an in-depth study of how such temporally modulated loss and gain can be modelled as a temporal effective medium as well as a detailed analysis of EM wave propagation inside both temporal multistepped and temporal effective metamaterials having loss and gain. It is worth noting that another set of interesting features in the temporal non-Hermitian PT symmetry system has been recently studied[55].

The schematic representation of the problem under study is shown in Fig. 1a. We consider a monochromatic EM wave traveling in a spatially unbounded medium modelled with a time-dependent relative permittivity $\varepsilon_r(t)$ (we will consider nonmagnetic materials having relative permeability $\mu_r = 1$). For times t < t₁, $\varepsilon_r$ of the whole medium is $\varepsilon_{r1} = \varepsilon'_{r1} - i\varepsilon''_{r1}$ with $\varepsilon''_{r1} < 0$ and $\varepsilon''_{r1} > 0$ denoting a medium with gain or loss, respectively. At $t = t_1$ the $\varepsilon_r$ of the entire medium is rapidly modified (approximately modelled mathematically as a step function) to $\varepsilon_{r2} = \varepsilon'_{r2} - i\varepsilon''_{r2}$ and then periodically alternated between this value and the original $\varepsilon_{r1} = \varepsilon'_{r1} - i\varepsilon''_{r1}$. As in [54], the total period of the temporal multistep is $\tau_{total} = \tau_1 + \tau_2 \ll T$, with *T* being the period of the incident signal and $\tau_{1,2}$ as the absolute time duration of each multistep. Finally, we can define the dimensionless temporal filling factors as $\Delta_{t1,2} = \tau_{1,2}/\tau_{total}$ (see Fig. 1a). Recently, we have shown in [54] that such temporal multistepped medium can be modeled by a temporally effective metamaterial having a $\varepsilon_r$ that is rapidly changed in time from $\varepsilon_{r1}$ to an effective $\varepsilon_{reff}$ at $t = t_1$. The effect of a temporally step function of $\varepsilon_r$ has been extensively studied in the past [38,39] demonstrating that a rapid change of $\varepsilon_r$ from one value $\varepsilon_{r1}$ to another $\varepsilon_{r2}$ (with a time duration $\ll T$) produce a temporal boundary which modifies the frequency of the medium from $f_1$ to $f_2 = (\sqrt{\varepsilon_{r1}}/\sqrt{\varepsilon_{r2}})f_1$ while keeps the wave number $\boldsymbol{k}$ unchanged. After implementing the Transfer Matrix method together with this condition we can arrive to the following analytical expression for the $\varepsilon_{reff}$ in temporal metamaterials[54]:

$$\varepsilon_{reff} = \frac{\varepsilon_{r1}\varepsilon_{r2}}{\varepsilon_{r1}\Delta_{t2}+\varepsilon_{r2}\Delta_{t1}} \quad (1)$$

Note the term $\Delta_{t1,2}$ is a dimensionless quantity. Now an interesting question may be raised: what will happen to the already propagating monochromatic EM wave if we consider a temporal multistepped metamaterial with gain and loss as in Fig. 1a? In the following we will answer this question by studying both numerically and theoretically (using Eq. (1)) the performance of the temporal multistepped metamaterial shown in Fig. 1a. It will be shown how the response of the equivalent temporally effective medium shown in Fig. 1b can be tuned depending on the induced loss and gain at each temporal step. It is important to note that, strictly speaking, the dispersion of the materials needs to be taken into account. However, if we consider that the material resonance frequencies are much larger than the frequency of operation, one can approximately assume the materials as dispersionless[52,56,57]. In this realm, we will also consider that only the imaginary part of relative permittivity, $\varepsilon_r''$, is modified to include either losses or gain. i.e., for the temporal multisteps from Fig. 1a the relative $\varepsilon_r$ will be changed from $\varepsilon_{r1} = \varepsilon_{r1}' - i\varepsilon_{r1}''$ to $\varepsilon_{r2} = \varepsilon_{r2}' - i\varepsilon_{r2}''$ at $t = t_1$ with the real parts stay unchanged, i.e., $\varepsilon_{r1}' = \varepsilon_{r2}' = \varepsilon_r'$ and then periodically changed the imaginary parts between $\varepsilon_{r2}''$ and $\varepsilon_{r1}''$.

With this setup, let us evaluate Eq. (1) using temporally periodic metamaterials with loss and gain. The theoretical results of the real and imaginary parts of the effective relative permittivity, i.e., $\varepsilon_{reff}'$ and $\varepsilon_{reff}''$, of the temporally effective medium as a function of $\varepsilon_{r1}''$ and $\varepsilon_{r2}''$ are shown in Fig. 1c,d, respectively, considering a filling factor for each temporal multistep of $\Delta_{t1,2} = 0.5$ and no change of the real component of $\varepsilon_r$ (from now onwards we will consider $\varepsilon_{r1}' = \varepsilon_{r2}' = \varepsilon_r' = 2$, the case of a small variation of $\varepsilon_{r1}'=1$ to $\varepsilon_{r2}'=1.1$ is shown as Supplementary Materials[58]). From Fig. 1c, one can notice how $\varepsilon_{reff}'$ has effectively negligible variations even when inducing temporal boundaries with large values of loss or gain. On the other hand, $\varepsilon_{reff}''$ (Fig. 1d) is strongly dependent of the loss and gain added into the system via the temporal multisteps between $\varepsilon_{r1}''$ and $\varepsilon_{r2}''$, as expected. From these results, one can create a temporal effective multistepped metamaterial having either a temporally effective loss or gain which can be tailored by correctly selecting the balance between $\varepsilon_{r1}''$ and $\varepsilon_{r2}''$. Moreover, as a temporal change of $\varepsilon_r$ modifies the frequency of the already propagating wave inside the material, the analytical results of the normalized frequency change ($\omega_{eff}/\omega_1 = (\omega_{eff}' + i\omega_{eff}'')/\omega_1$) are shown in the Supplementary Materials[58] for the same ranges of $\varepsilon_{r1}''$ and $\varepsilon_{r2}''$ shown in Fig. 1c,d.

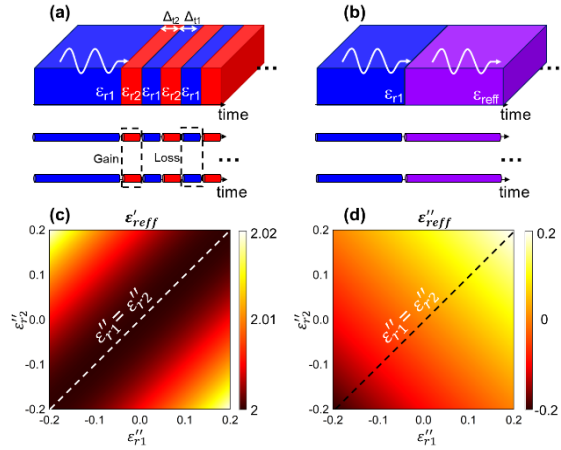

FIG. 1 (a) Schematic representation of the a multistepped temporal metamaterial made by alternating subperiods of gain and loss induced by $\varepsilon_r''$ along with its equivalent (b) temporally effective medium modelled as a single step change of $\varepsilon_r''$. The temporal filling factors $\Delta_{t1,2}$ are dimensionless quantities. Real and imaging parts of the effective relative permittivity (c) $\varepsilon_{reff}'$ and (d) $\varepsilon_{reff}''$, respectively, of the temporally effective medium shown in panel (b) calculated using Eq. (1) with $\varepsilon_{r1}' = \varepsilon_{r2}' = \varepsilon_r' = 2$ and $\Delta_{t1,2} = 0.5$.

Let us now evaluate the response of the proposed structures shown in Fig. 1a,b. Consider the temporal multistepped scenario shown in Fig. 1a where $\varepsilon_{r1}'' = 0$ for t < $t_1$ and it is periodically alternated between this value and $\varepsilon_{r2}'' = 0.02$ or $\varepsilon_{r2}'' = 0.1$ for t ≥ $t_1$. These temporal multistepped function of $\varepsilon_r$ are shown in Fig. 2a with the first panel representing the values of $\varepsilon_r'(t)$ while the blue curves from the middle and bottom panels are the $\varepsilon_r''(t)$ functions using $\varepsilon_{r2}'' = 0.02$ and $\varepsilon_{r2}'' = 0.1$, respectively. As in [54], we consider $\tau_{total} = 0.1T$ to fulfill the condition $\tau_{total} \ll T$, as described above. Now, using Eq. (1), such temporal multistepped metamaterials can be modeled each as a single step function of $\varepsilon_r$ where $\varepsilon_{r1}'' = 0$ for t < $t_1$ and then it is rapidly changed to $\varepsilon_{reff}'' = 0.01$ or $\varepsilon_{reff}'' = 0.05$, respectively, at t = $t_1$, while the real part of permittivity approximately stays the same. Both single step function of $\varepsilon_r$ are also plotted in the middle and bottom panels of Fig. 2a as green lines to guide the eye. As observed, both scenarios represent cases where there are no losses for t <

$t_1$ and then losses are added for t ≥ $t_1$. We evaluate both scenarios numerically using the transient solver of the commercial software COMSOL Multiphysics® with the same setup as in [54] by modelling the temporal changes of $\varepsilon_r''$ as equivalent changes of conductivity using periodic or single step analytical functions. We consider two-dimensional (2D) scenarios where a monochromatic plane wave with its electric field polarized along the *y* axis propagates along the *x* axis ($E_1 = \hat{y}e^{i(\omega_1 t - kx)}$) for t < $t_1$. The source is switched off when the changes of $\varepsilon_r$ are induced at t = $t_1$ = 37.2$T$ to better analyze how adding losses/gain in real time influence the EM wave already propagating in the medium.

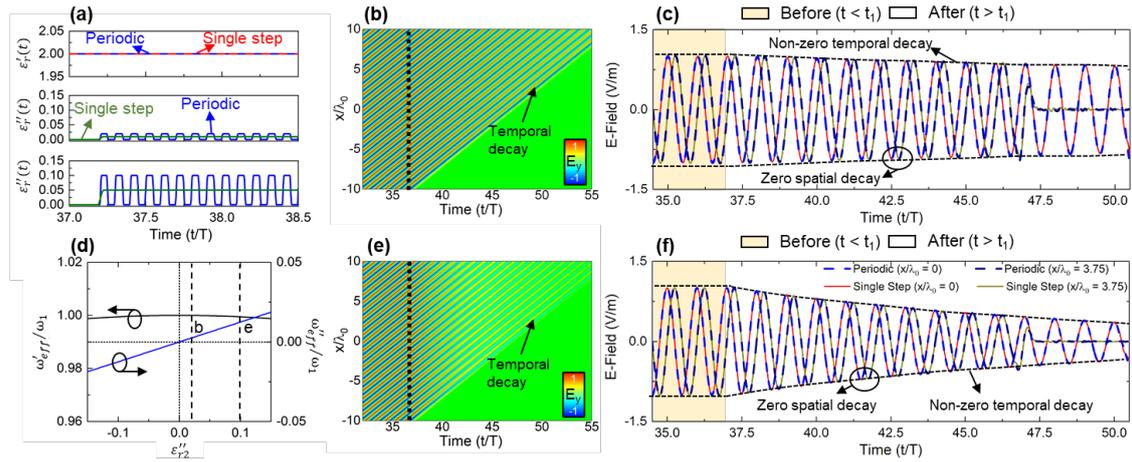

FIG. 2 (a) $\varepsilon_r(t)$ for the temporal multistepped metamaterials (blue) and their corresponding temporal effective media ($\varepsilon'_{reff}$, red and $\varepsilon''_{reff}$, green). We consider $\varepsilon'_{r1} = \varepsilon'_{r2} = \varepsilon'_r = 2$ stay unchanged, and only $\varepsilon''_{r2}$ (or $\varepsilon''_{reff}$) are modified in time. For the temporal multistepped media, $\varepsilon''_{r2} = 0$ for t < ($t_1$ = 37.2T) and then it is periodically modified between $\varepsilon''_{r2} = 0$ to either $\varepsilon''_{r2} = 0.02$ or $\varepsilon''_{r2} = 0.1$ for t ≥ $t_1$. The values of $\varepsilon''_{reff}$ are calculated using Eq. (1) with $\Delta_{t1,2}$ = 0.5. (b,e) Numerical results of the $E_y$ field distribution as a function of space and time for the scenarios shown in the middle and bottom panels of (a), respectively, using the temporal effective metamaterial modelled as a single step function of $\varepsilon_r''$. (c,f) $E_y$ field distribution extracted from (b,e), respectively, recorded at $x/\lambda_0 = 0$ and $x/\lambda_0 = 3.75$ using the temporal multistepped metamaterial (dashed) and temporal effective medium (solid). We show in (d) the effective $\omega'_{eff}/\omega_1$ (black) and $\omega''_{eff}/\omega_1$ (blue) when $\varepsilon_r''$ is changed from $\varepsilon''_{r1}$ to different values of $\varepsilon''_{r2}$ (to include loss or gain) at t = $t_1$ and then it is periodically alternated between $\varepsilon''_{r1}$ and $\varepsilon''_{r2}$.

The numerical results of the $E_y$ field distribution as a function of space ($x/\lambda_0$) and time for the single step functions of $\varepsilon_r$ from Fig. 2a (i.e., $\varepsilon''_{reff} = 0.01$ or $\varepsilon''_{reff} = 0.05$) are shown in Fig. 2b,e, respectively. From these figures, it is clearly seen how for t < ($t_1$ = 37.2$T$) the EM wave inside such media does not suffer any attenuation as a result of $\varepsilon''_{r1} = 0$, as expected. Now, at t = $t_1$ = 37.2$T$, $\varepsilon_r''(t)$ is temporally effectively modified to $\varepsilon''_{reff}$ with a single step function to introduce losses in the system. This will generate a complex change of frequency to $\omega_{eff}/\omega_1 = (\omega'_{eff} + i\omega''_{eff})/\omega_1$, as described before. The analytical results of $\omega_{eff}/\omega_1$ for the cases under study considering $\varepsilon''_{r1} = 0$ and multiple values of $\varepsilon''_{reff}$ are shown in Fig. 2d, for completeness. Note that, since $\varepsilon'_{r1} = \varepsilon'_{r2} =$ $\varepsilon'_r = 2$, the variation of the term $\omega'_{eff}/\omega_1$ is negligible and its value is always ~1 (for instance, it is 0.99999 and 0.99986 for the two cases under study in Fig. 2b,e, respectively, again demonstrating that the assumption of dispersionless scenario is justified for the proposed structures). On the other hand, the component $\omega''_{eff}/\omega_1$ will strongly depend on the value of $\varepsilon''_{reff}$. This can be observed in Fig. 2d where it is shown how $\omega''_{eff}/\omega_1$ is positive for both cases in Fig. 2b,e (represented as vertical dashed lines in Fig. 2d to guide the eye) but it is larger for the case with $\varepsilon''_{reff} = 0.05$ compared to when $\varepsilon''_{reff} = 0.01$, as expected. These parameters will influence the propagation of the EM wave already traveling in such media.

As shown in Fig. 2b,e, at t = $t_1$ the temporal boundary is introduced and the source is switched off for times t > ($t_1$ = 37.2$T$). As described above, a temporal boundary will create a set of a FW and a BW waves [38,39]. However, note that even when a BW wave is also produced in the results shown in Fig. 2b,e, only the FW wave is clearly visible for t > ($t_1$ = 37.2$T$) as the temporal change of $\varepsilon_r$ is too small ($\varepsilon'_r$ is not changed while $\varepsilon''_r$ is changed from $\varepsilon''_{r1} = 0$ to $\varepsilon''_{reff} = 0.05$ or $\varepsilon''_{reff} = 0.01$) and the BW is weak. Interestingly, also note how the wavelength of the waves in Fig. 2b,e is not changed for t > ($t_1$ = 37.2$T$), meaning that wave number ***k***, which is real-valued, is not modified, as expected[51]. However, an interesting physical phenomenon occurs: the amplitude of the EM wave decays in time in the entire space as it propagates along the *x* axis. This intriguing phenomenon can be attributed to the fact that the new normalized effective frequency $\omega_{eff}/\omega_1$ is complex and has an imaginary part different to zero. As $\omega''_{eff}/\omega_1$ is larger for the case with $\varepsilon''_{reff} = 0.05$, the amplitude of monochromatic EM wave will decay faster in time compared to the case with $\varepsilon''_{reff} = 0.01$ (see Fig. 2b,e). To better compare these results, we extracted the $E_y$ field distribution at single locations (*x/λ₀* =0 and *x/λ₀* =3.75) from Fig. 2b,e and the results are shown in Fig. 2c,f, respectively. Here we have included the $E_y$ field distribution considering the temporally multistepped metamaterial described by the temporal functions of $\varepsilon_r$ shown in Fig. 2a. These results corroborate how the amplitude of the EM wave decays in time, but not in space, as explained above, due to the fact that no losses were in the system for times t < ($t_1$ = 37.2$T$) hence only $\omega_{eff}/\omega_1$, but not ***k***, is complex for times t > ($t_1$ = 37.2$T$). As an aside, as expected, the results demonstrating how a temporally multistepped metamaterial with alternating losses (periodic and positive $\varepsilon''_r$) is in agreement with the model of a temporally effective medium with a single step function of $\varepsilon''_{reff}$. Animations showing the periodic and single step scenarios discussed in Fig. 2 are provided as Supplementary Materials[58].

In Fig. 2 we have discussed cases where no losses or gain were present before the change of $\varepsilon_r$ (i.e., $\varepsilon_{r1}'' = 0$). However, what would happen if losses were already present for t < t₁ and then we introduce temporal multistepped functions of $\varepsilon_r$ alternating between loss-loss or loss-gain? We discuss these cases in Fig. 3 where we consider that $\varepsilon_{r1}'' = 0.05$ with $\Delta_{t1,2} = 0.5$. First, let us consider a temporal multistepped metamaterial whose $\varepsilon_r''$ is $\varepsilon_{r1}'' = 0.05$ for t < t₁ and then it is periodically changing between $\varepsilon_{r1}'' = 0.05$ and $\varepsilon_{r2}'' = 0.01$, i.e., reducing losses. This temporal function of $\varepsilon_r''$ is shown in the top panel of Fig. 3a. By applying Eq. (1) to this temporal multistepped medium, the resulting equivalent temporal effective metamaterial is shown in the top panel of Fig. 3d where $\varepsilon_r''$ is rapidly changed from $\varepsilon_{r1}'' = 0.05$ to the effective value $\varepsilon_{reff}'' = 0.03$ in a single step. The numerical results of the spatial $E_y$ field distributions at different times before and after the change of $\varepsilon_r''$ are shown in Fig. 3a,d for the temporal multistepped and temporal effective medium, respectively. From these results it is clear how there is an excellent agreement between the results with the same field distribution for both temporal metamaterials. As the medium is lossy for t < t₁, the wavenumber $\boldsymbol{k}$ is a complex quantity, and thus the EM wave spatially decays along the propagation $x$ axis, as expected. Now, for t > t₁, the temporal effective medium has a value of $\varepsilon_{reff}'' = 0.03$, i.e. still lossy. As observed from Fig. 3a,d, the EM wave already traveling in the medium maintain the spatial decay (conservation of $\boldsymbol{k}$) and its maximum amplitude of 1 ($x/\lambda_0$ = -10) for times t < t₁ decays to ~0.63 at a time t/T = 50.1 when it propagates inside the temporal multistepped medium (Fig. 3a), in agreement with the equivalent temporal effective medium where the amplitude decays to ~0.64 .

Let us now introduce gain into the system. In Fig. 3b we use a temporal multistepped metamaterial with a value of $\varepsilon_{r1}'' = 0.05$ for t < t₁ and then it is periodically modified between $\varepsilon_{r1}'' = 0.05$ and $\varepsilon_{r2}'' = -0.05$ for t ≥ t₁. The equivalent temporal effective medium for this case is shown in Fig. 3e where, according to Eq. (1), $\varepsilon_r''$ is rapidly changed from $\varepsilon_{r1}'' = 0.05$ to $\varepsilon_{reff}'' = 0$ at t = t₁. Again, one can notice an excellent agreement between the results (Fig. 3b,e). Interestingly, note how the amplitude of the $E_y$ field is kept approximately the same while the EM propagates in the medium. For instance, it can be seen how the maximum amplitude of the wave is 1 (at $x/\lambda_0$ = -10) for times t < t₁ and, after inducing the temporal boundaries and propagating, it still is approximately the same amplitude of 1 at a time t/T = 50.1. As shown in Fig. 3e, following Eq. (1) the temporal multistepped metamaterial can be modelled as a temporal effective medium with $\varepsilon_{reff}'' = 0$, i.e., no losses for times t ≥ $t_1$. In this context, even when the EM waves inside the temporal metamaterials decay in space, these losses are compensated by the temporal modulation of $\varepsilon_r''$ with loss and gain.

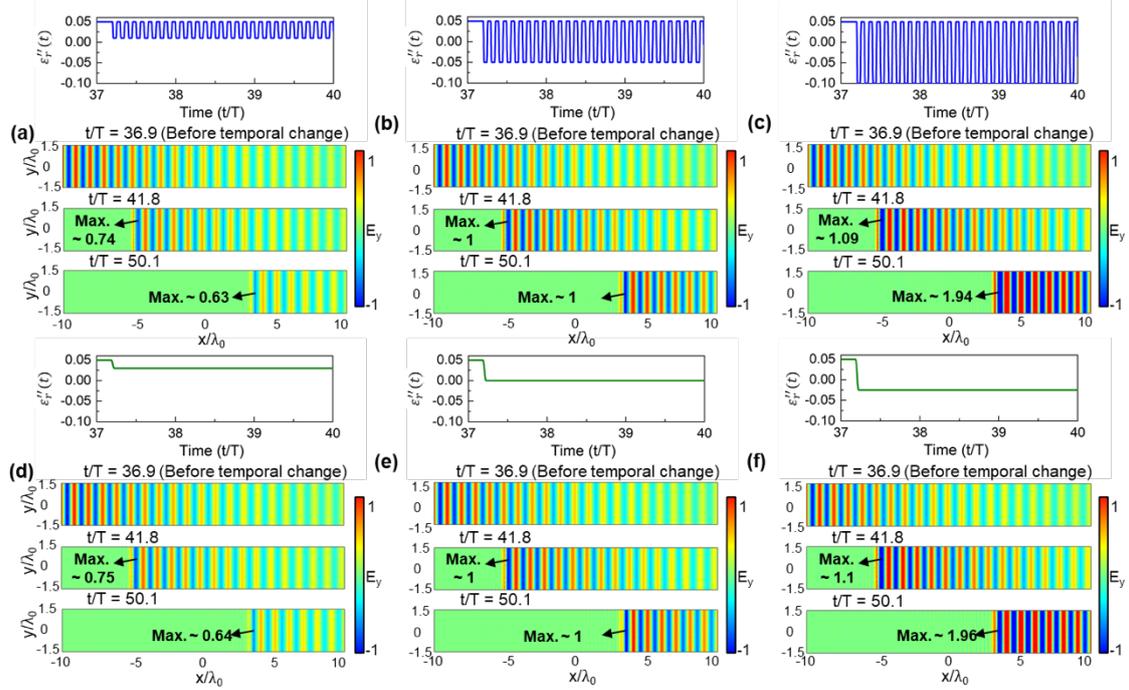

FIG. 3 Spatial distributions of the $E_y$ field at different times using (a-c) temporal multistepped metamaterials with $\varepsilon_{r1}'' = 0.05$ for t < ($t_1 = 37.2T$) and then periodically alternated between $\varepsilon_{r1}''= 0.05$ and: (a) $\varepsilon_{r2}''$=0.01, (b) $\varepsilon_{r2}''$=-0.05 and (c) $\varepsilon_{r2}''$=-0.1. (d-f) Spatial distribution of the $E_y$ field at different times using the corresponding temporal effective medium for cases shown in panels (a-c) respectively, modelled as single step functions. The temporal functions of $\varepsilon_r''(t)$ for each case are shown on the top panel for each temporal metamaterial.

What would happen if we increase the gain for each temporal multistepped metamaterial? We evaluate this scenario in Fig. 3c were we show the results of a temporal multistepped metamaterial with $\varepsilon_{r1}'' = 0.05$ for t < $t_1$ that is periodically modified between $\varepsilon_{r1}'' = 0.05$ and $\varepsilon_{r2}'' = -0.1$ for t ≥ $t_1$. The results of the temporal effective medium for this scenario are shown in Fig. 3f using a single step function with $\varepsilon_{reff}'' = -0.025$ (calculated from Eq. (1)). As observed, as $\varepsilon_{reff}'' < 0$, the temporal effective metamaterial behaves as a medium with gain meaning that the amplitude of the $E_y$ field in the whole space will increase in time as the wave propagates for t ≥ $t_1$. As in Fig. 3b,e, this temporal response compensates the spatial losses due to $\varepsilon_{r1}''$ being $\varepsilon_{r1}'' = 0.05$ for t < $t_1$. This can be corroborated in Fig. 3c,f where it is observed how the maximum amplitude of 1 (at $x/\lambda_0$ = -10) for times t < $t_1$ is increased up to almost twice its value for a time t/T = 50.1. It is important to note that, as discussed in Fig. 2, all temporal boundaries induced in Fig. 3 will also produce a BW wave but this is not visible due to the small change of $\varepsilon_r$. This BW wave could be further increased if one use a temporally effective metamaterial with a larger value of gain (i.e. $\varepsilon_{reff}''$<<0). In this context, both FW and BW waves will increase their amplitude in time in a similar fashion to those results shown in Fig. 3c,f (not shown here). For completeness, we provide the values of the instantaneous $E_y$ field distribution as a function of time for all the cases from Fig. 3 at two different positions along the $x$ axis in the Supplementary Materials along with animations showing the $E_y$ field distribution in the whole space for the same cases[58].

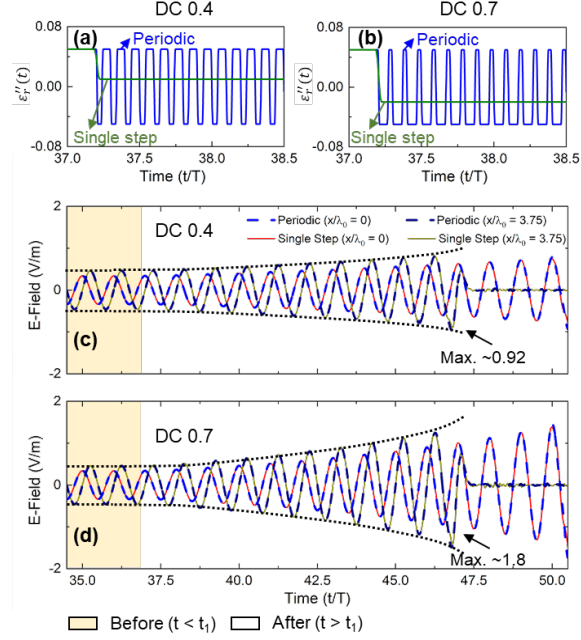

FIG. 4 (a,b) $\varepsilon_r''(t)$ for temporal multistepped metamaterials (blue) and their corresponding temporal effective medium (green). The temporal multistepped metamaterial is changed from $\varepsilon_{r1}'' = 0.05$ and then periodically alternated between $\varepsilon_r'' = 0.05$ and $\varepsilon_r''=-0.05$ for $t \geq t_1$. The temporal effective media are obtained from Eq. (1) considering a DC of (a) 0.4 and (b) 0.7. (c,d) $E_y$ field distribution for the cases shown in panels (a,b), respectively, calculated at $x/\lambda_0 = 0$ and $x/\lambda_0 = 3.75$ using temporal multistepped metamaterial (dashed) and temporal effective medium (solid) having loss and gain.

Finally, in Fig. 1-3 we have discussed different temporal metamaterials with loss and gain considering equal time duration for each temporal step (i.e., $\Delta_{t1,2} = 0.5$). As a final study, in Fig. 4 we present the numerical results of the $E_y$ field distribution recorded at $x/\lambda_0 = 0$ and $x/\lambda_0 = 3.75$ for the temporal multistepped metamaterial shown in Fig. 3b ($\varepsilon_{r1}'' = 0.05$ for t < $t_1$ and then it is periodically modified between $\varepsilon_{r1}'' = 0.05$ and $\varepsilon_{r2}'' = -0.05$ for $t \geq t_1$) but now using a different duty cycle (DC) for the duration of the temporal multisteps: DC of 0.4 and 0.7. The $\varepsilon_r''$ for these temporal multistepped metamaterials are shown in Fig. 4a,b along with their equivalent functions of $\varepsilon_{reff}''$ calculated from Eq. (1) ($\varepsilon_{reff}'' = 0.01$ and $\varepsilon_{reff}'' = -0.02$, respectively). As observed, the spatial decay is not compensated by the periodic temporal multisteps with loss and gain for the DC of 0.4, producing a decay of the maximum amplitude of the wave from 1 to ~0.92 (measured at $x/\lambda_0 = 3.75$ in Fig. 4c). However, for a DC of 0.7, the temporal effective medium generates a material with gain ($\varepsilon''_{reff} = -0.02$) which in turn produce a temporal amplification of the wave across the entire space, compensating the spatial decay (note that the maximum amplitude of the wave is increased in Fig. 4d). These results demonstrate how, changing the DC of a temporal multistepped metamaterial with loss and gain can be tuned to generate a desired temporal effective medium modelled as a single step.

In summary, we have investigated the performance of temporal multistepped metamaterials with loss and gain showing how such media can be equivalently modelled by a temporal effective permittivity following a single step function in time having an effective loss or gain behavior. We have studied several scenarios including temporal metamaterials inducing loss-to-loss, loss-to-no-loss, and loss-to-gain multisteps showing how an EM wave traveling inside such medium can experience spatiotemporal decay or amplification depending on the values of loss and gain added into the system. The physics behind such temporal modulation of media has been explored in detail and evaluated both numerically and theoretically, demonstrating good agreement between the results. We envision that these results may open new avenues in the exploration of wave-matter interactions by temporally modulating the EM properties of media for potential applications in loss reduction and signal amplification while keeping the wavenumber unchanged.

V.P.-P. acknowledges support from Newcastle University (Newcastle University Research Fellowship). N.E. would like to acknowledge the partial support from the Simons Foundation/Collaboration on Symmetry-Driven Extreme Wave Phenomena.

varying media: A generalization of the Kramers-Kronig relations. *Phys. Rev. B* **103**, 144303 (2021).
58. Materials and methods are available as supplementary materials.

# Supplementary materials:
# Temporal metamaterials with gain and loss

*Victor Pacheco-Peña[1] and Nader Engheta[2]*

[1]*School of Mathematics, Statistics and Physics, Newcastle University, Newcastle Upon Tyne, NE1 7RU, United Kingdom*
[2]*Department of Electrical and Systems Engineering, University of Pennsylvania, Philadelphia, PA 19104, USA*
*email: Victor.Pacheco-Pena@newcastle.ac.uk, engheta@ee.upenn.edu

**Table of content:**

1. Plots for $\omega'_{eff}/\omega_1$ and $\omega''_{eff}/\omega_1$.

2. Temporal variation between zero losses and losses.

3. Small variations of $\varepsilon'_r$

4. $E_y$ field: changing $\varepsilon''_{r2}$ between lossy-to-lossy and lossy-to-gain.

5. List of supplementary movies.

# 1. Plots for $\omega'_{eff}/\omega_1$ and $\omega''_{eff}/\omega_1$.

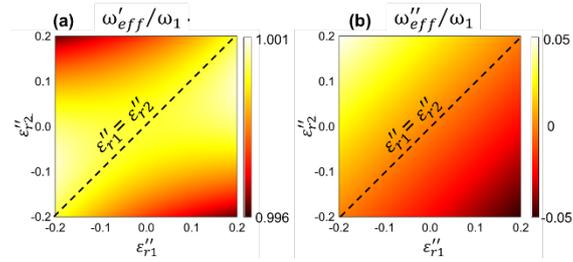

FIG. S1 (a) $\omega'_{eff}/\omega_1$ and (b) $\omega''_{eff}/\omega_1$ values for the temporally effective medium shown in Fig. 1b of the main text calculated using Eq. (1) of the main manuscript using $\varepsilon'_{r1} = \varepsilon'_{r2} = \varepsilon'_r = 2$ and $\Delta_{t1,2} = 0.5$.

# 2. Temporal variation of electric field when the material changes temporally between the case of zero loss and the case of non-zero loss.

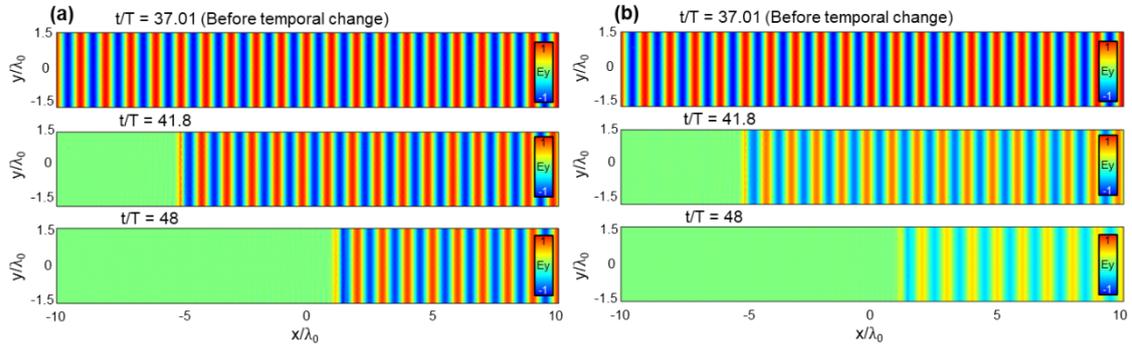

FIG. S2 Spatial distributions of the $E_y$ field at different times evaluated using the temporal multistepped metamaterials from Fig. 2 of the main with $\varepsilon''_{r1} = 0$ for t < ($t_1$ = 37.2$T$) and then periodically alternated between $\varepsilon''_{r1}= 0$ and: (a) $\varepsilon''_{r2}$=0.02 and (b) $\varepsilon''_{r2}$=0.1.

## 3. Small temporal variations of $\varepsilon'_r$

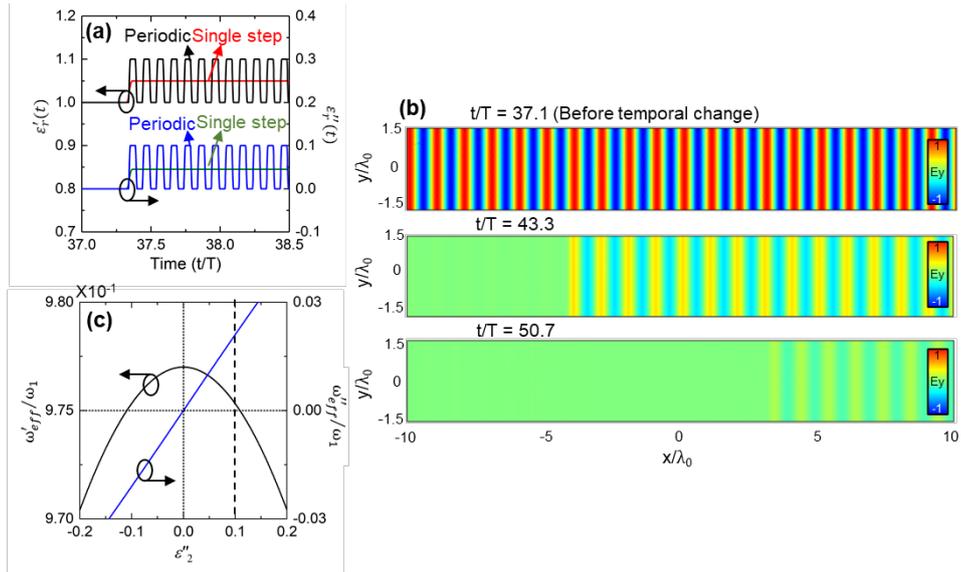

FIG. S3 Numerical and analytical results showing the performance of a temporal multistepped metamaterial with loss and gain and its corresponding temporal effective medium when $\varepsilon'_r$ is not kept constant in time and it is $\varepsilon'_{r1} = 1$ for for t < ($t_1$ = 37.2$T$) and then it is periodically changed between $\varepsilon'_{r1} = 1$ and $\varepsilon'_{r1} = 1.1$. Here the $\varepsilon''_r$ is modified as in the case discussed in Fig. 2e,f of the main text with $\varepsilon''_{r1} = 0$ and $\varepsilon''_{r2} = 0.1$. The values for the real and imaginary terms of the complex $\varepsilon_r$ for the temporal multistepped metamaterial along with its temporal effective medium modelled as a single step are shown in (a). (b) Effective frequency and (c) spatial $E_y$ field distribution at different times for the temporal multistepped metamaterial with temporal losses.

## 4. E$_y$ field: changing $\varepsilon''_{r2}$ between lossy-to-lossy and lossy-to-gain cases.

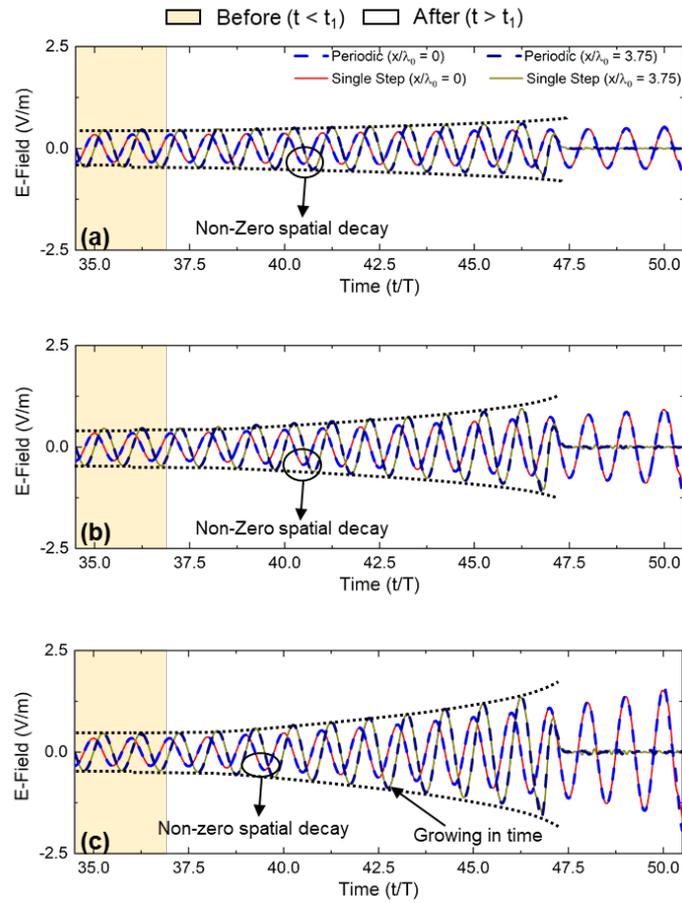

FIG. S4 $E_y$ field distribution as a function of time (evaluated at $x/\lambda_0 = 0$ and $x/\lambda_0 = 3.75$) for the cases discussed in Fig. 3 of the main text considering: (a) $\varepsilon''_{r2}$=0.01, (b) $\varepsilon''_{r2}$=-0.05 and (c) $\varepsilon''_{r2}$=-0.1.

5. List of supplementary movies

- **Supplementary movie 1:** Spatial $E_y$ field distribution for the temporal multistepped metamaterial and its equivalent temporal effective medium considering the case shown in Fig. 2b,c of the main text with $\varepsilon'_{r1} = \varepsilon'_{r2} = \varepsilon'_r = 2$, $\varepsilon''_{r2} = 0$ for $t < t_1$ and then it is periodically changed between $\varepsilon''_{r2} = 0$ and $\varepsilon''_{r2} = 0.02$ for $t \geq t_1$.

- **Supplementary movie 2:** Spatial $E_y$ field distribution for the temporal multistepped metamaterial and its equivalent temporal effective medium considering the case shown in Fig. 2e,f of the main text with $\varepsilon'_{r1} = \varepsilon'_{r2} = \varepsilon'_r = 2$, $\varepsilon''_{r2} = 0$ for $t < t_1$ and then it is periodically changed between $\varepsilon''_{r2} = 0$ and $\varepsilon''_{r2} = 0.1$ for $t \geq t_1$.

- **Supplementary movie 3:** Spatial $E_y$ field distribution for the temporal multistepped metamaterial and its equivalent temporal effective medium considering the case shown in Fig. 3a,d of the main text with $\varepsilon'_{r1} = \varepsilon'_{r2} = \varepsilon'_r = 2$, $\varepsilon''_{r2} = 0.05$ for $t < t_1$ and then it is periodically changed between $\varepsilon''_{r2} = 0.05$ and $\varepsilon''_{r2} = 0.01$ for $t \geq t_1$.

- **Supplementary movie 4:** Spatial $E_y$ field distribution for the temporal multistepped metamaterial and its equivalent temporal effective medium considering the case shown in Fig. 3b,e of the main text with $\varepsilon'_{r1} = \varepsilon'_{r2} = \varepsilon'_r = 2$, $\varepsilon''_{r2} = 0.05$ for $t < t_1$ and then it is periodically changed between $\varepsilon''_{r2} = 0.05$ and $\varepsilon''_{r2} = -0.05$ for $t \geq t_1$.

- **Supplementary movie 5:** Spatial $E_y$ field distribution for the temporal multistepped metamaterial and its equivalent temporal effective medium considering the case shown in Fig. 3c,f of the main text with $\varepsilon'_{r1} = \varepsilon'_{r2} = \varepsilon'_r = 2$, $\varepsilon''_{r2} = 0.05$ for $t < t_1$ and then it is periodically changed between $\varepsilon''_{r2} = 0.05$ and $\varepsilon''_{r2} = -0.1$ for $t \geq t_1$.